\documentclass[aps,prx,twocolumn,superscriptaddress,showpacs,amsmath,amssymb]{revtex4-1}

\usepackage{amsfonts}
\usepackage{amssymb}
\usepackage{graphicx}
\usepackage{color}

\usepackage{bm}
\usepackage{braket}

\usepackage{hyperref}

\definecolor{dblue}{rgb}{0.0, 0.0, 0.5}

\definecolor{dgreen}{rgb}{0.0, 0.5, 0.0}

\usepackage[normalem]{ulem}

\begin{document}

\title{Fractional corner magnetization of collinear antiferromagnets}

\author{Haruki Watanabe} \email[]{hwatanabe@g.ecc.u-tokyo.ac.jp}
\affiliation{Department of
Applied Physics, University of Tokyo, Tokyo 113-8656, Japan.}

\author{Yasuyuki Kato} \email[]{yasuyuki.kato@ap.t.u-tokyo.ac.jp}
\affiliation{Department of
Applied Physics, University of Tokyo, Tokyo 113-8656, Japan.}

\author{Hoi Chun Po} \email[]{hcpo@mit.edu}
\affiliation{Department of Physics, Massachusetts Institute of Technology, Cambridge, Massachusetts 02139, USA.}

\author{Yukitoshi Motome} \email[]{motome@ap.t.u-tokyo.ac.jp}
\affiliation{Department of
Applied Physics, University of Tokyo, Tokyo 113-8656, Japan.}

\begin{abstract}
Recent studies revealed that the electric multipole moments of insulators result in fractional electric charges localized to the hinges and corners of the sample.
We here explore the magnetic analog of this relation. We show that a collinear antiferromagnet with spin $S$ defined on a $d$-dimensional cubic lattice features fractionally quantized magnetization $M_{\text{c}}^z=S/2^d$ at the corners. 
We find that the quantization is robust even in the presence of gapless excitations originating from the spontaneous formation of the N\'eel order, although the localization length diverges, suggesting a power-law localization of the corner magnetization.
When the spin rotational symmetry about the $z$ axis is explicitly broken, the corner magnetization is no longer sharply quantized. Even in this case, we numerically find that the deviation from the quantized value is negligibly small based on quantum Monte Carlo simulations.
\end{abstract}

\maketitle

\section{Introduction}
Multipole insulators feature fractional electric charges bound to hinges and corners of the system~\cite{benalcazarScience,benalcazar2017}.
Previous studies clarified the relation between the electric multiple moments of insulators and the boundary charges~\cite{benalcazar2017,Luka2020,PhysRevB.102.165120,Vanderbilt2020}.
It was recently shown that ionic crystals may exhibit a fractional corner charge reflecting their octupole moment, despite the fact that they are trivial in that they do not feature protected ground-state  quantum entanglement~\cite{2009.04845}.

Given that fractional corner charges can appear in ionic crystals, it is natural to expect fractional ``corner magnetizations" in the magnetic analog of ionic crystals. In collinear antiferromagnets, up and down spins can be respectively regarded as positively and negatively charged ions (of the charge associated with the conserved spin rotation symmetry). Indeed, a recent work discussed the magnetic analog of the Benalcazar-Bernevig-Hughes model~\cite{2010.05402}. 
However, two fundamental problems inherit to spin systems that are absent in the electric counterparts have not been addressed in the earlier works: (i) The U(1) symmetry underlying the charge conservation (i.e., the spin rotational symmetry about the $z$ axis) may be explicitly broken because of the crystal anisotropy and spin-orbit coupling, and (ii) excitations may not be gapped because the spontaneous formation of the N\'eel order results in gapless Nambu-Goldstone excitations.  For these reasons, the results for electric multiple moments established in the previous works~\cite{benalcazar2017,benalcazar2018,Luka2020,PhysRevB.102.165120,Vanderbilt2020,2009.04845} cannot be directly applied to spin systems.

In this work, we address these problems by numerically studying the antiferromagnetic Heisenberg model.  We find that, despite the above two issues, the fractional corner magnetizations indeed appear in the collinear antiferromagnets. When the spin U(1) symmetry is exact, the corner magnetization is quantized to $S/2^d$ in the antiferromagnetic ground state of spin-$S$ models defined on a $d$-dimensional cubic lattice, even when the bulk N\'eel order is not fully saturated due to quantum fluctuations. In contrast to electric multipole insulators where corner charges are exponentially localized, 
the localization length of the corner magnetization diverges in the gapless limit of the isotropic Heisenberg model due to the gapless spin wave excitations.
As expected from the symmetry-protected nature of the phase, the corner spin quantization is lost when the U(1) symmetry is not exact. Yet, for symmetry broken cases by the bond anisotropy we studied, we find the deviation from the ideal value  to be negligibly small as compared to $S/2^d$.

This work is organized as follows. 
As a warm-up, we first study the edge magnetization of one-dimensional Heisenberg models and discuss its relation to the bulk polarization in Sec.~\ref{sec1D}.
We then move on to the corner magnetization of two- and three-dimensional Heisenberg models in Sec.~\ref{sec2D}.
Finally we discuss material candidates and conclude in Sec.~\ref{conclusion}.

\begin{figure*}
\begin{center}
\includegraphics[width=0.75\textwidth]{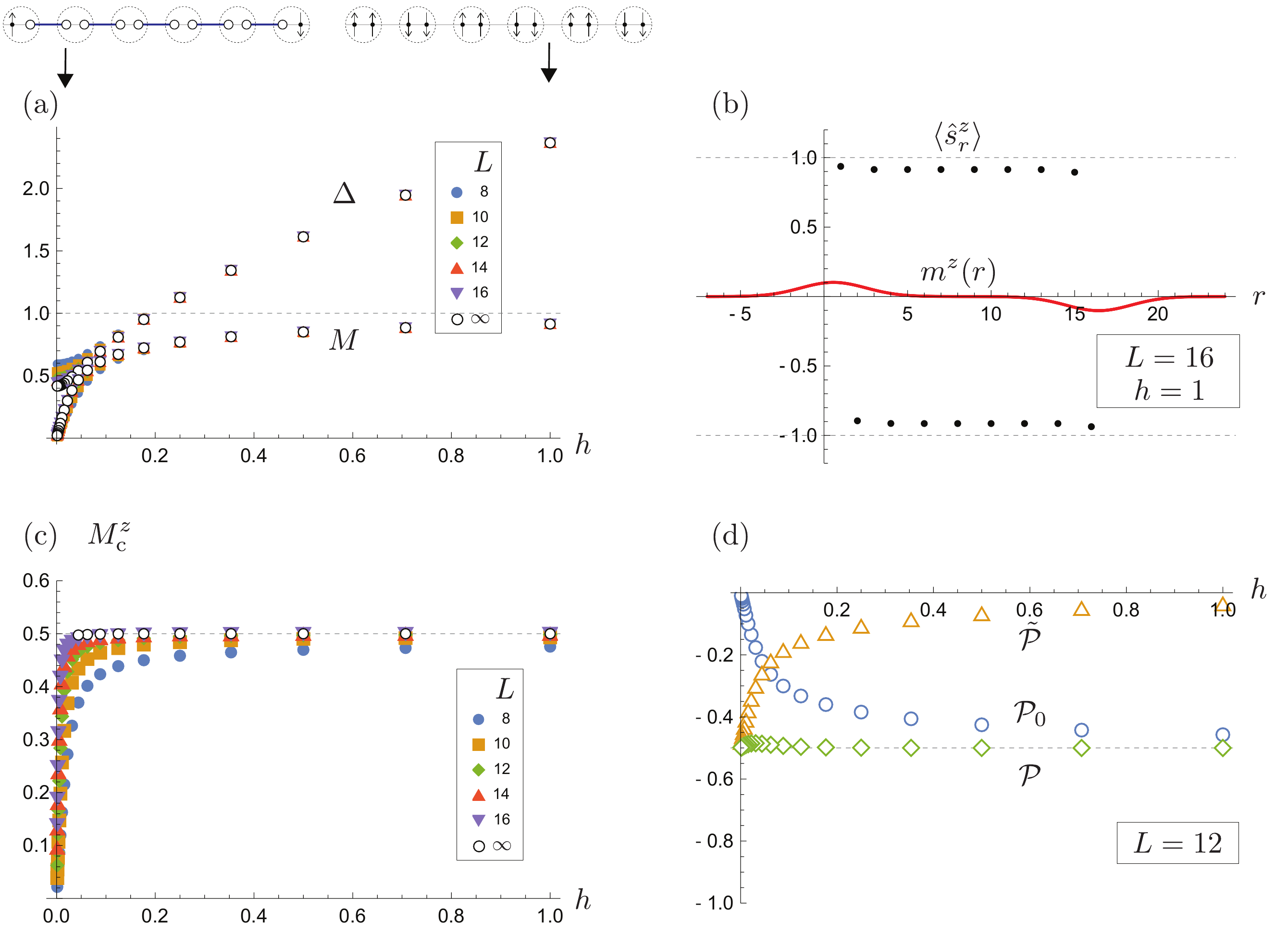}
\caption{\label{fig1DS1} 
Exact diagonalization of the $S=1$ Heisenberg chain.
(a) The excitation gap $\Delta$ and the staggered magnetization $M$ under the PBC as functions of the staggered magnetic field $h$. The $L=\infty$ values are extrapolated using the data for $L=8,10,12,14,$ and $16$. 
The insets shows the schematic figures of the ground state for $h=0$ (the gapped Haldane state) and $h=\infty$ (classical N\'eel ordered state); the arrows and the small circles connected by the lines represent the $S=1/2$ degrees of freedom decomposed from the original $S=1$ and the singlet pairs, respectively.
(b) The expectation value of the $z$-spin component at site $r$, $\langle \hat{s}_r^z\rangle$, and the coarse-grained magnetization $m^z(r)$ [Eq.~\eqref{cgmag}] under the OBC for $L=16$ and $h=1$.
(c) The edge magnetization $M_{\text{c}}^z$ [Eq.~\eqref{edgeM}] under the OBC as a function of $h$. The extrapolation to $L\to\infty$ fails for $h<0.04$.
(d) The classical polarization $\mathcal{P}_0$ [Eq.~\eqref{classicalP}], the Berry phase correction $\tilde{\mathcal{P}}$ [Eq.~\eqref{BPcorr}], and the sum $\mathcal{P}=\tilde{\mathcal{P}}+\mathcal{P}_0$ for $L=12$.
}
\end{center}
\end{figure*}

\section{One dimension}
\label{sec1D}
Let us first discuss the Heisenberg spin chain as a canonical example of spin models featuring the edge magnetization.  The Hamiltonian under the open boundary condition (OBC) reads
\begin{equation}
\hat{H}=J\sum_{r=1}^{L-1}\hat{\bm{s}}_{r}\cdot\hat{\bm{s}}_{r+1}+h\sum_{r=1}^L(-1)^r\hat{s}_{r}^z,
\label{H1DOBC}
\end{equation}
where $\hat{\bm{s}}_r = (\hat{s}_r^x, \hat{s}_r^y, \hat{s}_r^z)$ is the spin-$S$ operator at site $r$. 
The first term describes the nearest-neighbor antiferromagnetic interaction and the second term is the Zeeman coupling to a staggered magnetic field along the $z$ direction. 
We introduce the staggered field in order to realize a magnetic analog of the ionic crystals; note that the N\'eel order is absent when $h=0$ since continuous symmetries cannot be broken spontaneously in one dimension. 
We also consider this model under the periodic boundary condition (PBC) by adding the term $J \hat{\bm{s}}_{L}\cdot\hat{\bm{s}}_{1}$ to $\hat{H}$.  In this section, the system size $L$ is assumed to be an even integer to match the number of two sublattices.  For brevity, the coupling constant $J$ and the lattice constant are set to be unity.

\subsection{Spin-1 model}
\label{sec1DS1}
We start with the $S=1$ model.  When $h=0$, the system is in the Haldane phase, which is a symmetry-protected topological phase (with the bulk excitation gap $\Delta\simeq0.4105$
~\cite{PhysRevB.48.3844}) protected by either one of the following three symmetries: the time-reversal symmetry, the $\mathbb{Z}_2\times\mathbb{Z}_2$ spin rotational symmetry, or the spatial inversion symmetry about the bond center
~\cite{PhysRevB.85.075125}.
Under the OBC, each edge supports an emergent $S=1/2$ degrees of freedom and the ground state is fourfold degenerate in the thermodynamic limit of large $L$ [see the left inset of Fig.~\ref{fig1DS1}(a)].  

The staggered field $h>0$ breaks all the protecting symmetries of the Haldane phase and trivializes the system. 
As $h$ is increased from $0$ to $+\infty$, the system develops the forced N\'eel order and the ground state is smoothly connected to the product state limit $\langle\hat{s}_r^z\rangle=(-1)^{r-1}$ [the right inset of Fig.~\ref{fig1DS1}(a)] without closing the bulk gap $\Delta$~\cite{PhysRevLett.114.177204}, as demonstrated by the exact diagonalization up to $L=16$ in Fig.~\ref{fig1DS1}(a). The $L=\infty$ value $\Delta(\infty)$ is obtained by the system-size extrapolation by using the data for $L=8$, $10$, $12$, $14$, and $16$ with the linear fitting of $\log[\Delta(L)-\Delta(\infty)]$. The $L=\infty$ values of other quantities (e.g., $M$ and $M_{\text{c}}^z$ introduced below) are determined in the same way.

\begin{figure*}
\begin{center}
\includegraphics[width=0.75\textwidth]{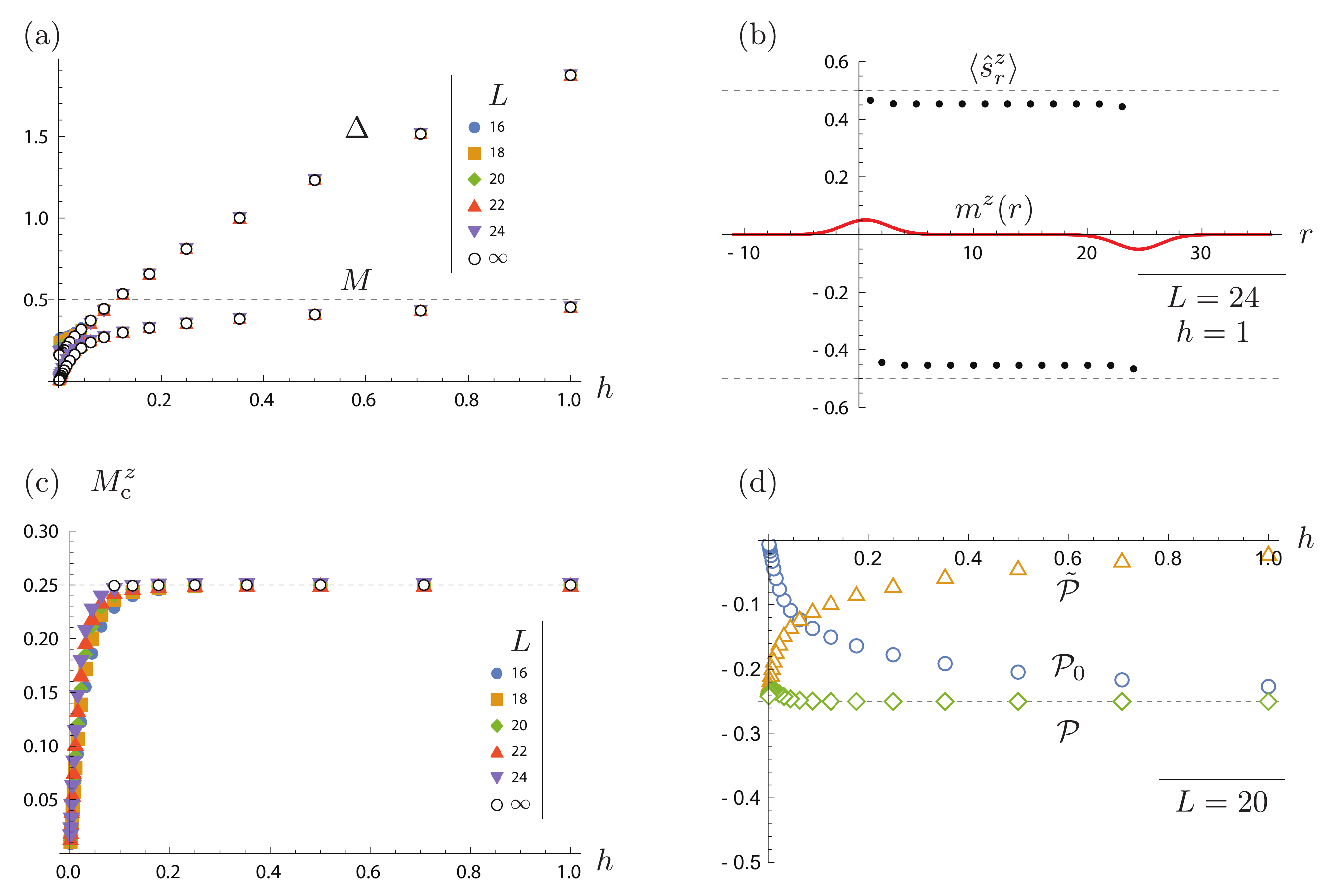}
\caption{\label{fig1DS12} 
The same as Fig.~\ref{fig1DS1} but for the $S=1/2$ model.
(a) The excitation gap $\Delta$ and the staggered magnetization $M$ under the PBC as functions of the staggered magnetic field $h$. The $L=\infty$ values are extrapolated using the data for $L=16,18,20,22,$ and $24$.
(b) The expectation value $\langle \hat{s}_r^z\rangle$ and the coarse-grained magnetization $m^z(r)$ [Eq.~\eqref{cgmag}] under the OBC for $L=24$ and $h=1$.
(c) The edge magnetization $M_{\text{c}}^z$ [Eq.~\eqref{edgeM}] under the OBC as a function of $h$. The extrapolation to $L\to\infty$ fails for $h<0.07$.
(d) The classical polarization $\mathcal{P}_0$ [Eq.~\eqref{classicalP}], the Berry phase correction $\tilde{\mathcal{P}}$ [Eq.~\eqref{BPcorr}], and the sum $\mathcal{P}=\tilde{\mathcal{P}}+\mathcal{P}_0$ for $L=20$..
}
\end{center}
\end{figure*}

To study the edge magnetization of a finite-$L$ sample, we compute the local magnetization $\langle \hat{s}_r^z\rangle$ on each site using the unique ground state under the OBC selected by $h>0$. We then derive the coarse-grained magnetization $m^z(r)$ defined by the convolution integral~\cite{jackson2007classical}
\begin{equation}
m^z(r)\equiv \sum_{r'=1}^Lg(r-r')\langle \hat{s}_{r'}^z\rangle
\label{cgmag}
\end{equation}
with the Gaussian $g(r)=(2\pi \lambda^2)^{-1/2}e^{-r^2/(2\lambda^2)}$. This smoothed magnetization has a maximum and a minimum near the edges and vanishes deep inside the bulk as far as $\lambda$ is chosen sufficiently large. In this work, we set $\lambda=2$ in all calculations (Appendix \ref{app:lambda}). The results are plotted by the red curve in Fig.~\ref{fig1DS1}(b) together with $\langle \hat{s}_r^z\rangle$ for $L=16$ and $h=1$.  We define the magnetization localized to the left edge by
\begin{equation}
M_{\text{c}}^z\equiv\int_{-\infty}^{(L+1)/2}dr\,m^z(r).
\label{edgeM}
\end{equation}
The results while varying $h$ are shown in Fig.~\ref{fig1DS1}(c), suggesting that the edge magnetization $M_{\text{c}}^z$ is quantized to $+1/2$ in the limit of large $L$ for any $h>0$. This is in sharp contrast to the bulk N\'eel order parameter 
\begin{equation}
M\equiv \frac{1}{L}\sum_{r=1}^L(-1)^{r-1}\langle \hat{s}_{r}^z\rangle
\end{equation}
that does not reach the saturation value $M=1$ for any finite $h$ [see Fig.~\ref{fig1DS1}(a)]. 
The fractional edge magnetization is a reminiscent of the $S=1/2$ edge mode at $h=0$: the $S_z=\pm 1/2$ levels of the edge spin split as a result of the applied staggered field and give the saturation magnetization in the ground state.

This result has a simple interpretation in terms of the bulk polarization.  When the spin rotational symmetry about the $z$ axis is exact, one can define the bulk polarization $\mathcal{P}$ with respect to the conserved U(1) charge $\hat{S}^z=\sum_{r=1}^L\hat{s}_r^z$, which implies the appearance of the ``surface charge" (i.e., the edge magnetization) $M_{\text{c}}^z=\mathcal{P}$ (mod 1) when the OBC is imposed.  The polarization $\mathcal{P}$ is given by
\begin{equation}
\mathcal{P} = \mathcal{P}_0 + \tilde{\mathcal{P}},
\end{equation}
where
\begin{equation}
\mathcal{P}_0=\frac{1}{L}\sum_{r=1}^Lr\langle\hat{s}_r^z\rangle\,\,\,\left(=-\frac{1}{2}M\right)
\label{classicalP}
\end{equation}
is the classical contribution evaluated under the PBC and 
\begin{align}
\tilde{\mathcal{P}}=\int_0^{2\pi}\frac{d\theta}{2\pi}i\langle \theta|\partial_\theta |\theta\rangle\quad\text{mod }1,
\label{BPcorr}
\end{align}
is the Berry phase correction~\cite{PhysRevB.77.094431}. The state $|\theta\rangle$ in Eq.~\eqref{BPcorr} is the ground state under the twisted boundary condition, under which the term $J(e^{-i\theta}\hat{s}_L^+\hat{s}_1^-+\text{h.c.})/2+\hat{s}_L^z\hat{s}_1^z$ is added to $\hat{H}$ in Eq.~\eqref{H1DOBC}~\footnote{In the actual numerical calculation, instead of directly evaluating Eq.~\eqref{BPcorr} that requires a smooth gauge fixing of $|\theta\rangle$, we used the discretized version of the Berry phase formula derived in Appendix C of Ref.~\cite{PhysRevB.78.054431} with $\Delta\theta=2\pi/100$. Also, see Appendix~\ref{appResta} for an alternative formulation of the bulk polarization.}. 
The spatial inversion symmetry about a site (which remains a symmetry even under $h>0$) quantizes the polarization $\mathcal{P}$ to either $0$ or $1/2$ (mod 1)  regardless of the value of $h$. Figure~\ref{fig1DS1}(d) numerically confirms the quantization of $\mathcal{P}$ to $1/2$ in our present model and supports the fractional edge magnetization $M_{\text{c}}^z=1/2$ shown in Fig.~\ref{fig1DS1}(c)~\footnote{The small deviation of $\mathcal{P}$ from $1/2$ in Fig.~\ref{fig1DS1}(d) is a finite-size effect originating from the $\theta$ dependence of $\langle \theta|\hat{P}_0|\theta\rangle$. If $\langle \hat{P}_0\rangle$ in $\mathcal{P}$ is replaced by $\int_0^{2\pi}d\theta/(2\pi) \langle \theta|\hat{P}_0|\theta\rangle$, the quantization of $\mathcal{P}$ becomes exact even for a finite $L$~\cite{PhysRevX.8.021065}.}.

\subsection{Spin-1/2 model}
The edge magnetization $M_{\text{c}}^z=\pm 1/2$ in the $S=1$ model may not be surprising because it can be understood simply by the emergent $S=1/2$ edge degrees of freedom in the Haldane phase, which polarizes readily for any staggered field $h>0$. Here we instead discuss the $S=1/2$ Heisenberg model and show that the edge magnetization is $M_{\text{c}}^z=\pm 1/4$ in this case. This example clarifies that the existence of a topological counterpart is not a prerequisite for  
the emergence of fractional edge magnetizations.

To this end, we perform exactly the same calculations for the $S=1/2$ model as in Sec.~\ref{sec1DS1}. 
We used $L=16, 18, 20, 22$, and $24$ for the extrapolation to $L\to \infty$  for the $S=1/2$ case.  Our results, summarized in Fig.~\ref{fig1DS12}, are qualitatively the same as the $S=1$ case, although the values of the bulk polarization $\mathcal{P}$ and the edge magnetization $M_{\text{c}}^z$ are halved as compared to the $S=1$ case because the effective charge unit becomes $1/2$.  Note that the apparent nonzero gap $\Delta$ for $h=0$ in Fig.~\ref{fig1DS12}(a) is an artifact of the current extrapolation scheme that assumes an exponential decay as a function of the system size $L$. If we fit the gap assuming a power-law decay instead, the extrapolated value becomes negligibly small.

\begin{figure}
\begin{center}
\includegraphics[width=0.8\columnwidth]{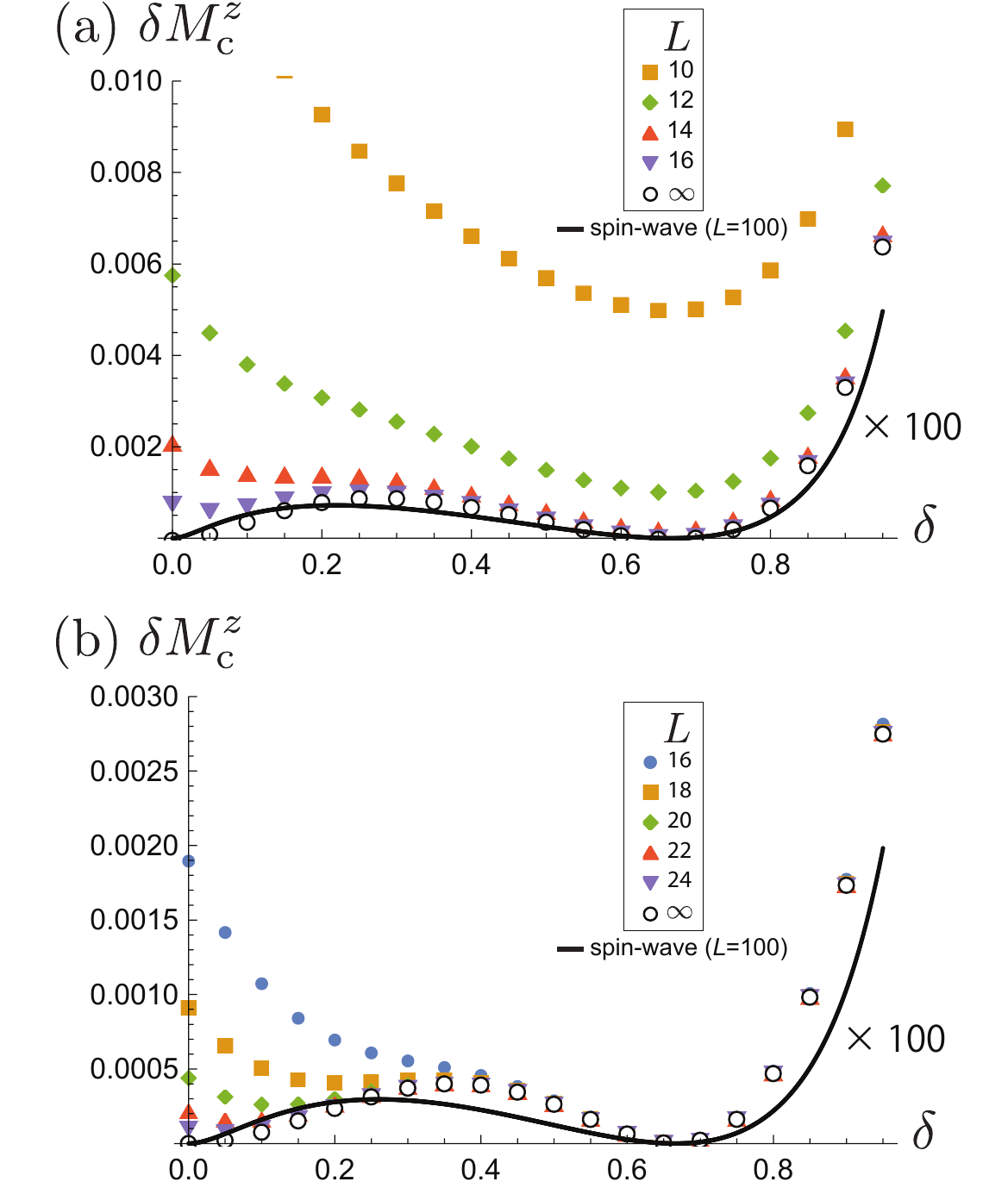}
\caption{\label{fig1Da} 
Deviation of the edge magnetization, $\delta M_{\text{c}}^z\equiv S/2-M_{\text{c}}^z$, as a function of $\delta$ at $h=0.25$ for the (a) $S=1$ and (b) $S=1/2$ models. 
For comparison, results from the linear spin-wave approximation are plotted by black lines.
The spin-wave calculation captures the correct qualitative behavior but significantly underestimate the size of the deviation, and as such
the absolute values of the spin-wave results are multiplied by a factor of $100$.}
\end{center}
\end{figure}

\subsection{Effect of anisotropy}
Now let us discuss the effect of the U(1) symmetry breaking.
As an example, here we consider an additional term to Eq.~\eqref{H1DOBC}  given by
\begin{align}
\hat{H}'=-\delta\sum_{r=1}^L(2\hat{s}_{r}^x\hat{s}_{r+1}^x+\hat{s}_{r}^y\hat{s}_{r+1}^y),
\end{align}
with $0\leq\delta<1$.  The spin rotational symmetry about the $z$ axis is explicitly broken when $\delta\neq0$ or $2/3$.  When $\delta=2/3$, $J_x=-J_y$ $(=-1/3)$ and the U(1) symmetry can be restored by applying the $\pi$-rotation about the $x$ axis for spins on only one of the two sublattices. In this case, the ground state is given by the ferromagnetic state in the rotated basis, which corresponds to the classical N\'eel state [i.e., $\langle\hat{s}_r^z\rangle=(-1)^{r-1}$] in the original basis.  

Since the exact U(1) symmetry was the key for the argument supporting the quantized edge magnetization based on the bulk polarization, the explicit breaking of the U(1) symmetry may, in principle, completely destroy the edge magnetization. To see if this is the case, we compute the deviation of the edge magnetization from $S/2$ as a function of $\delta$. The result for $h=0.25$ is shown in Fig.~\ref{fig1Da}(a) for the $S=1$ model and Fig.~\ref{fig1Da}(c) for the $S=1/2$ model. The $L=\infty$ values are determined by the same extrapolation procedure as discussed above.  Although the edge magnetization is no longer quantized to $S/2$ except for $\delta=0$ and $2/3$,  the deviation of $M_{\text{c}}^z$ turns out to be about $1$\% even for a fairly large $\delta\sim 1$.  As expected, $\delta M_{\text{c}}^z$ vanishes in the $L=\infty$ limit at the U(1) symmetric points $\delta=0$ and  $2/3$. For a finite $L$, $\delta M_{\text{c}}^z$ at $\delta=2/3$ are generally smaller than $\delta M_{\text{c}}^z$ at $\delta=0$. This is because nonzero $\delta$ tends to increase the bulk excitation gap.

\begin{figure*}
\begin{center}
\includegraphics[width=1\textwidth ,trim =0 1150 0 0, clip]{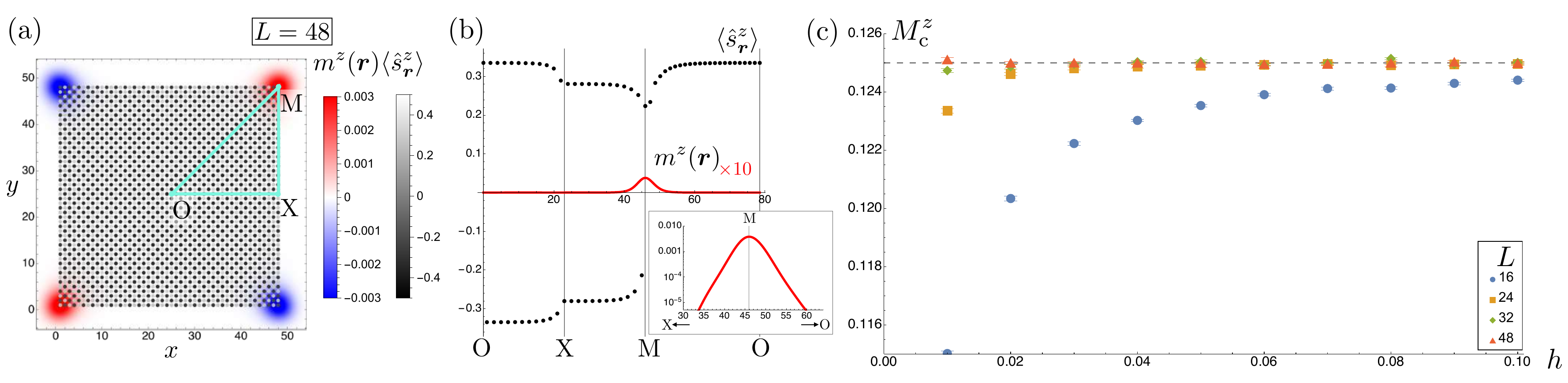}
\caption{
\label{fig2D_1} 
QMC simulations of the $S=1/2$ Heisenberg model on a square lattice under the OBC.
(a,b) Real-space distribution of the expectation value $\langle \hat{s}_{\bm{r}}^z\rangle_{L,h}$ and the coarse-grained magnetization $m_{L,h}^z(\bm{r})$ [Eq.~\eqref{smoothenmLhz}] under the OBC for 
 $L=\beta=48$, $\delta=0$, and $h=0.01$. 
(b) shows $\langle \hat{s}_{\bm{r}}^z\rangle_{L,h}$ and $m_{L,h}^z(\bm{r})$ along the 
lines connecting the representative points 
in (a): ${\rm O}=(L/2+1,L/2+1)$, ${\rm X}=(L-1,L/2+1)$, and ${\rm M}=(L,L)$.
The inset shows $m_{L,h}^z({\bf r})$ near M in the logarithmic scale, indicating that $m_{L,h}^z(\bm{r})$ is exponentially localized to corners. 
(c) The edge magnetization $(M_{\text{c}}^z)_{L,h}$ [Eq.~\eqref{edgeMhighD}] as a function of $h$. 
The inverse temperature is set to be $\beta = L$.
The statistical errors in the QMC data are smaller than the symbol sizes.
}
\end{center}
\end{figure*}

We also perform the linear spin-wave calculation. We first represent the spin operators using creation/annihilation operators of bosons assuming the classical N\'eel order, then linearize the Hamiltonian by dropping all interacting terms, and diagonalize the Hamiltonian following the algorithm summarized in Ref.~\cite{COLPA1978327}. The results for $L=100$ are plotted in Figs.~\ref{fig1Da}(b) and \ref{fig1Da}(d) for the $S=1$ and $1/2$ cases, respectively.  The spin wave results qualitatively agree with those by the exact diagonalization, although the absolute values are much smaller in the spin wave approach, which tends to underestimate quantum fluctuations.

\section{Higher dimensions}
\label{sec2D}
Let us move on to higher dimensions.  Our numerical calculations in this section are based on quantum Monte Carlo (QMC) simulations.
\subsection{Isotropic case}
\label{isotropic}
We consider the antiferromagnetic Heisenberg model on the $d$-dimensional ($d=2$ or $3$) cubic lattice
\begin{align}
\hat{H}=J\sum_{\langle\bm{r},\bm{r}'\rangle}\hat{\bm{s}}_{\bm{r}}\cdot\hat{\bm{s}}_{\bm{r}'},
\label{HhDOBC}
\end{align}
where $\langle\bm{r},\bm{r}'\rangle$ represents the nearest neighboring sites. We set $J=1$ as before.   
Unlike the one-dimensional model, in dimensions higher than one, the SO(3) spin rotational symmetry of the Hamiltonian is spontaneously broken down to the U(1) rotation about the direction of the N\'eel ordered moment at zero temperature.
This spontaneously broken symmetry gives rise to gapless Nambu-Goldstone excitations,  invalidating the direct application of the results~\cite{benalcazar2017,benalcazar2018,Luka2020,PhysRevB.102.165120,2010.05402,Vanderbilt2020,2009.04845} for gapped systems. Nevertheless, we show in the following that the corner magnetization of antiferromagnets is still quantized to $M_{\text{c}}^z=S/2^{d+1}$ despite the gapless nature of the phase.

To discuss the corner magnetization, we want to impose the OBC with a finite linear dimension $L$ as we did in the previous section.  As is well known, however, the true ground state of any finite-size system respects all symmetries of the Hamiltonian. For the Heisenberg model, this is guaranteed by the Marshall-Lieb-Mattis theorem~\cite{doi:10.1098/rspa.1955.0200,doi:10.1063/1.1724276} stating that the ground state has the lowest possible total spin, i.e., spin 0 when $L$ is even and spin $1/2$ when $L$ is odd.

The standard trick to overcome this difficulty is to temporary introduce the staggered field $h$ by adding
\begin{align}
h\sum_{\bm{r}}e^{i\bm{Q}\cdot\bm{r}}\hat{s}_{\bm{r}}^z
\end{align}
to the Hamiltonian in Eq.~\eqref{H1DOBC}, where $\bm{Q}=(\pi,\pi)$ for $d=2$ and $\bm{Q}=(\pi,\pi,\pi)$ for $d=3$. The applied field breaks the SO(3) symmetry of  the Hamiltonian and induces a gap $\Delta\propto \sqrt{h}$ to the Nambu-Goldstone excitations. 
Therefore, the system is effectively gapped at every stage of the calculation. There is an alternative approach based on the long-range property of correlation functions as we discuss in Sec.~\ref{LRO}.

Given $L$ and $h>0$, the definition of  the corner magnetization  $M_{\text{c}}^z$ in two and three dimensions is the direct extension of that of the edge magnetization in one dimension.
We first compute the expectation value $\langle \hat{s}_{\bm{r}}^z\rangle_{L,h}$ using the ground state with $h>0$ and then introduce the smoothened magnetization by the Gaussian convolution
\begin{align}
m_{L,h}^z(\bm{r})=\sum_{\bm{r}'}\,g(\bm{r}-\bm{r}')\langle \hat{s}_{\bm{r}'}^z\rangle_{L,h},
\label{smoothenmLhz}
\end{align}
where $g(\bm{r})=(2\pi \lambda^2)^{-d/2}e^{-|\bm{r}|^2/(2\lambda^2)}$ with $\lambda=2$. We define the corner magnetization by the integral of $m_{L,h}^z({\bm r})$ over the corner region $R$:
\begin{align}
(M_{\text{c}}^z)_{L,h}=\int_{R}d^dr\,m_{L,h}^z({\bm r})\label{edgeMhighD}.
\end{align}
The specific choice of $R$ is not important as far as it fully contains a single corner, since the smoothened magnetization $m_{L,h}^z(\bm{r})$ is nonzero only near corners of the system as shown below. 
In our calculation we use $-\infty<x,y\leq (L+1)/2$ in two dimensions and $-\infty<x,y,z\leq (L+1)/2$ in three dimensions. Finally, we switch off the staggered field \emph{after} taking the thermodynamic limit:
\begin{align}
M_{\text{c}}^z\equiv\lim_{h\rightarrow+0}\lim_{L\rightarrow\infty}(M_{\text{c}}^z)_{L,h}.\label{Mczlim1}
\end{align}

\begin{figure*}
\begin{center}
\includegraphics[width=1\textwidth,trim= 0 1120 0 0,clip]{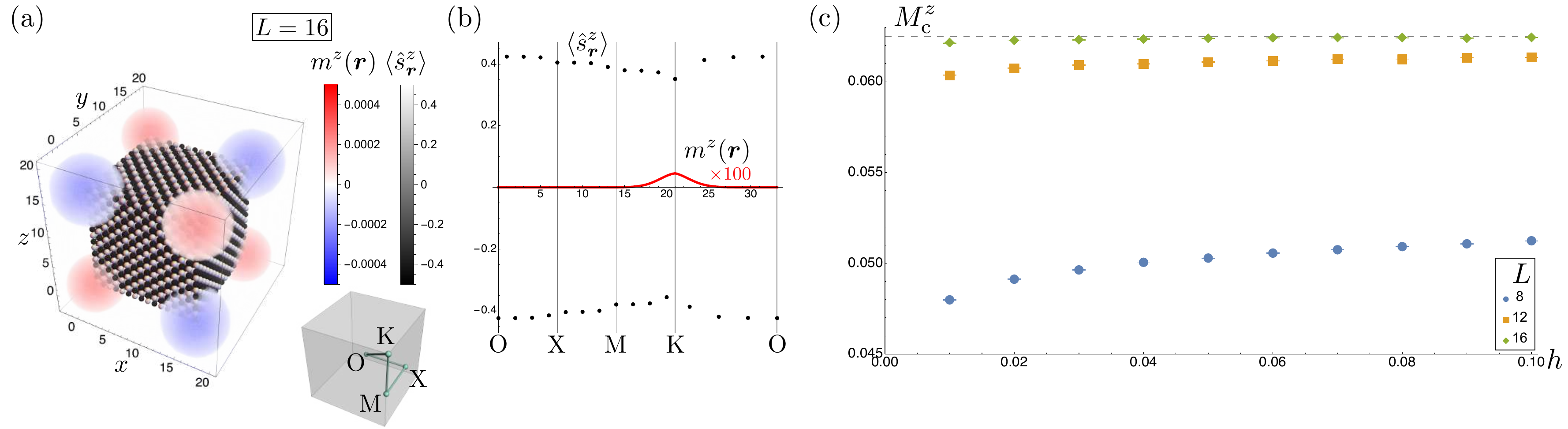}
\caption{
\label{fig3D_1} 
QMC simulations of the $S=1/2$ Heisenberg model on a simple cubic lattice under the OBC.
(a,b) Real space distribution of the expectation value $\langle \hat{s}_{\bm{r}}^z\rangle$ and the coarse-grained magnetization $m^z(\bm{r})$ [Eq.~\eqref{smoothenmLhz}] under the OBC for $L=\beta/2=16$, $\delta=0$ and $h=0.01$. 
The lattice points are labeled by $\bm{r}=(x,y,z)$ with three integers $1 \leq x,y,z \leq L$.
The representative points are defined as 
${\rm O}=(L/2+1,L/2,L/2+1)$, 
${\rm X}=(L,L/2,L/2+1)$, 
${\rm M}=(L,1,L/2+1)$
and
${\rm K}=(L,1,L)$.
Plot (b) shows $\langle \hat{s}_{\bm{r}}^z\rangle$ and $m^z(\bm{r})$ along the straight lines connecting the representative points 
(${\rm O}\to{\rm X}\to{\rm M}\to{\rm K}\to{\rm O}$) shown in the inset of (a).
(c) The edge magnetization $M_{\text{c}}^z$ [Eq.~\eqref{edgeMhighD}] as a function of $h$. 
The inverse temperature is set to be $\beta = 2L$.
The statistical errors in the QMC data are smaller than the symbols.
}
\end{center}
\end{figure*}

Similar to the polarization argument for the one-dimensional case in Sec.~\ref{sec1DS1}, one can compute the corner magnetization under the field, $(M_{\text{c}}^z)_{L,h}$ in Eq.~\eqref{edgeMhighD}, in several ways.  One simple approach is known as  ``filling anomaly"~\cite{benalcazar2018}, which gives the fractional part of the corner charge in terms of the total U(1) charge $Q$ in the system and the number of point-group related corners $N$.  The U(1) charge $Q$ in our present problem is given by the eigenvalue of the total magnetization operator 
\begin{align}
\hat{S}_{\text{tot}}^z = \sum_{\bm{r}} \hat{s}_{\bm{r}}^z.
\end{align}
For example, if we impose the OBC with odd $L$ for two-dimensional square lattice systems, the fourfold rotational symmetry 
$C_4$ about the center site of the system gives $N=4$.  Furthermore, $Q=\pm S$ mod $4e$ depending on the direction of the center spin, where $e$ is the unit of the U(1) charge defined by
\begin{equation}
e\equiv\begin{cases}
1&(S=0,1,2,\cdots)\\
1/2&(S=1/2, 3/2,\cdots)
\end{cases}.
\end{equation}
Therefore, 
\begin{align}
(M_{\text{c}}^z)_{L,h}=\frac{Q}{N}=\pm \frac{S}{4}\mod e.
\label{fa2}
\end{align}
For a three-dimensional cube, $N=8$ because of the additional inversion symmetry $I$ about the center. 
Viewing the three-dimensional system as stacked two-dimensional layers with alternating corner magnetizations $\pm S/4$, we find $Q=\pm S$ mod $2e$. (This ambiguity of $Q$ can be understood from the coordination number of each lattice site. See Ref.~\cite{2009.04845} for more details.)  The sign corresponds to the corner magnetization of the topmost (or the bottommost) layer.
Hence,
\begin{align}
(M_{\text{c}}^z)_{L,h}=\frac{Q}{N}=\pm \frac{S}{8}\mod \frac{e}{4}.
\label{fa3}
\end{align}
Note that the filling anomaly formulas in Eqs.~\eqref{fa2} and \eqref{fa3} assume that the excess charge in the system is localized to corners; i.e., the bulk, surfaces, and hinges are all charge neutral. 
Furthermore, the linear dimension of the system, $L$, must be an odd integer; otherwise the point-group symmetry is broken and the total U(1) charge $Q$ vanishes.  However, since the corner magnetization is a local property determined by the configuration of $\langle\hat{s}_{\bm{r}}^z\rangle$ near a corner,  the value of $(M_{\text{c}}^z)_{L,h}$ should be unchanged for even $L$ when $L$ is sufficiently large.
In fact, the calculation of the corner charge based on the bulk multipole moment~\cite{PhysRevB.102.165120} is free from the parity of $L$. 

Both rotation $C_4$ and inversion $I$ are the symmetries of the system even in the presence of the staggered field $h$ and the symmetry quantization of $(M_{\text{c}}^z)_{L,h}$ remains effective. Therefore, from Eq.~\eqref{Mczlim1}, we conclude
\begin{align}
M_{\text{c}}^z\equiv\lim_{h\rightarrow+0}\lim_{L\rightarrow\infty}(M_{\text{c}}^z)_{L,h}=\pm \frac{S}{2^d}
\label{Mcz23D}
\end{align}
modulo $e$ for $d=2$ and $e/4$ for $d=3$ in the antiferromagnetic phase.

To verify the picture above, we perform unbiased QMC calculations based on Feynman's path integral~\cite{doi:10.1143/JPSJ.73.1379} for the $S=1/2$ models in two and three dimensions.
To update worldline configurations, we adopt a modified version~\cite{PhysRevE.79.021104} of the directed-loop algorithm~\cite{PhysRevE.66.046701}.
For each parameter set, we perform typically $10^5$ Monte Carlo sweeps for the thermalization and the measurement in 128 independent Markov chains.
In the QMC simulations, the local magnetization $\langle \hat{\bm s}_{\bm r} \rangle_{L,h}$ is given by the Gibbs ensemble average with the sufficiently large inverse temperature $\beta$ instead of the ground state expectation value. 
The estimate is exact within the statistical error. 
Given $\langle \hat{\bm s}_{\bm r} \rangle_{L,h}$, we compute the smoothened magnetization $m_{L,h}^z(\bm{r})$ and the corner magnetization $(M_{\text{c}}^z)_{L,h}$ using Eqs.~\eqref{smoothenmLhz} and \eqref{edgeMhighD}.

Our results are summarized in Fig.~\ref{fig2D_1} for the two-dimensional model and Fig.~\ref{fig3D_1}  for the three-dimensional model.  As shown in the panels (a) and (b) of Figs.~\ref{fig2D_1} and \ref{fig3D_1}, the smoothened magnetization $m_{L,h}^z(\bm{r})$ is nonzero only near the corners of the systems, which justifies our definition of the corner magnetization as an integral of $m_{L,h}^z(\bm{r})$ over a corner region as  in Eq.~\eqref{edgeMhighD}.  Figures~\ref{fig2D_1}(c) and ~\ref{fig3D_1}(c) illustrate how the corner magnetization $(M_{\text{c}}^z)_{L,h}$ approaches the ideal value $S/2^d$ as the system size $L$ is increased while the SO(3) symmetry is broken by a nonzero $h$, verifying our conclusion in Eq.~\eqref{Mcz23D}. These plots also demonstrate the importance of the order of the two limits, $L\to \infty$ and $h\to +0$: if the order is reversed, one finds $\lim_{L\rightarrow\infty}\lim_{h\rightarrow+0}(M_{\text{c}}^z)_{L,h}=0$.

\begin{figure}
\begin{center}
\includegraphics[width=1\columnwidth]{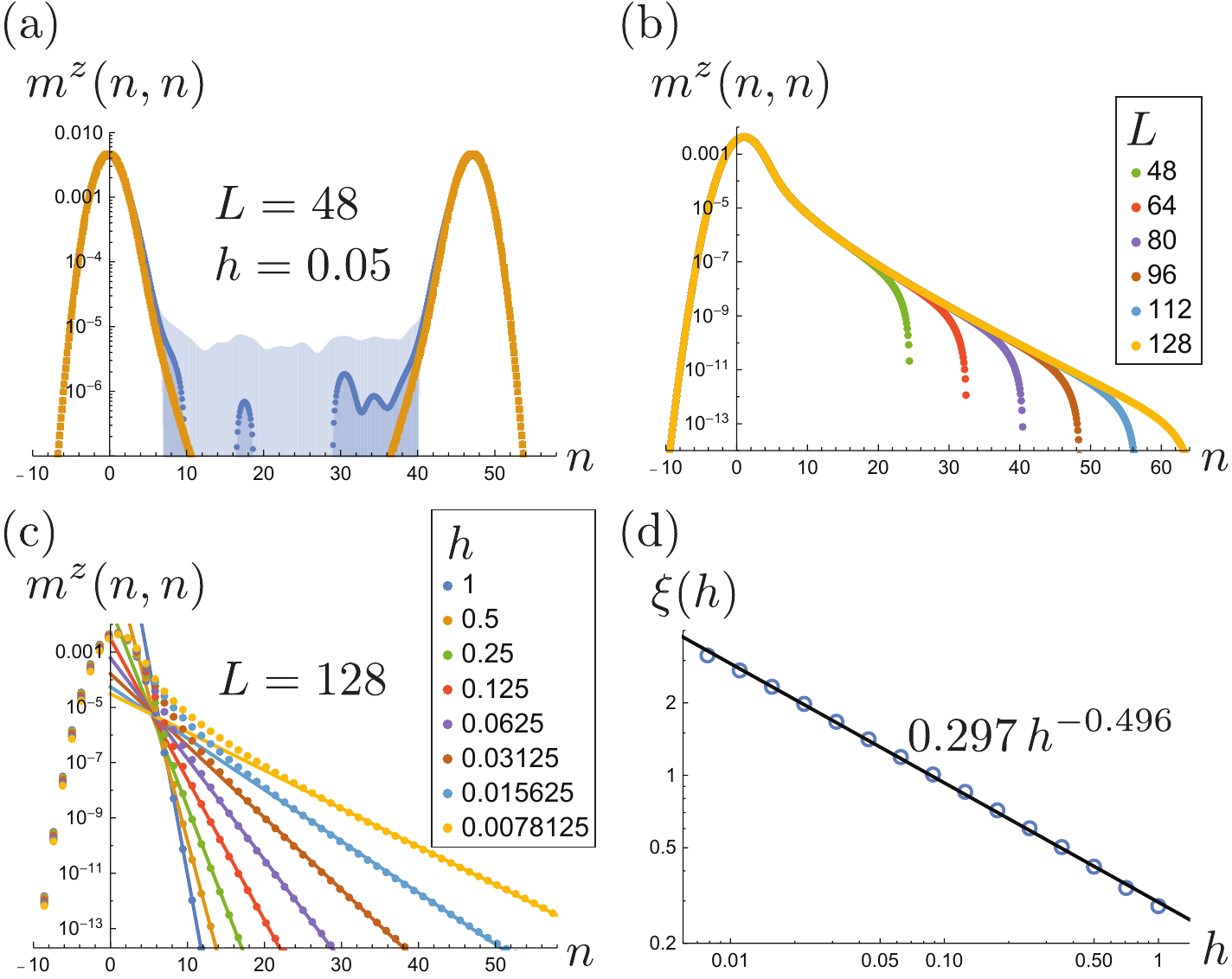}
\caption{
(a) The coarse-grained magnetization $m_{L,h}^z(\bm{r})$ along the diagonal line $\bm{r}=(n,n)$ ($n\in\mathbb{N}$) based on the spin-wave theory (orange) and the QMC simulation (blue). The QMC data for $|m_{L,h}^z(\bm{r})|\lesssim10^{-5}$ is buried within the statistical error. 
(b) The spin-wave results of $m_{L,h}^z(n,n)$ under the staggered field $h=2^{-7}$ ($=0.0078125$), suggesting that $L=128$ is sufficiently large for $n<55$ and $h\geq2^{-7}$. 
(c) The spin-wave results of $m_{L,h}^z(n,n)$ for $L=128$. The localization length $\xi(h)$ of the corner charge is determined by the slope of the linear fitting line.
(d) The $h$ dependence of the localization length $\xi(h)$ in (c). The slope of the fitting line (black) is $0.496$.
\label{fig2D_3} 
}
\end{center}
\end{figure}

The exponential localization of $m_{L,h}^z(\bm{r})$ found in Figs.~\ref{fig2D_1} and \ref{fig3D_1} may be understood from the bulk excitation gap $\Delta\propto \sqrt{h}$ induced by the staggered field $h>0$. This implies that the localization length of the corner magnetization diverges in the limit of $h\rightarrow+0$.  To confirm this understanding, we systematically study the $h$ dependence of $m_{L,h}^z(\bm{r})$ along the diagonal line $\bm{r}=(n,n)$ in the two-dimensional case. Figure~\ref{fig2D_3}(a) compares the QMC results to the linear spin-wave ones at $h=0.05$ and $L=48$. While the two results agree well with each other, the QMC results for $8\lesssim n\lesssim 40$  severely suffer from the statistical error. Figure~\ref{fig2D_3}(b) shows the $L$ dependence of the spatial decay obtained by the spin-wave approximation, suggesting that $L$ must be fairly large to avoid the finite-size effect. For these reasons, we use the results from the spin-wave theory for $L=128$ for this analysis.  Figure~\ref{fig2D_3}(c) suggests that $m_{L,h}^z(n,n)$ decays exponentially for sufficiently large $n$ for each $h>0$. We determine the localization length $\xi(h)$ from the slope of the fitting line. The $h$ dependence of the localization length is plotted in Fig.~\ref{fig2D_3}(d), which implies a power-law divergence $\propto h^{-a}$ $(a=0.496)$.  This behavior is consistent with the $h$ dependence of the excitation gap $\Delta\propto\sqrt{h}$.

Our result suggests that $m^z(\bm{r})$ in the $h\rightarrow+0$ limit exhibits only a power-law decay $\propto r^{-b}$ due to the presence of gapless spin-wave excitations.  This does not invalidate the quantization of the corner magnetization $M_c^z=S/2^d$ as long as the correct order of limit is assumed and the corner region $R$ is chosen much larger than $\xi(h)^d$ for each $h$.  For the convergence of the corner magnetization $M_c^z=\int_Rd^{d}r\,m^z(\bm{r})$, which is guaranteed by the upper bound of the integral $S/2^{d}$ set by the filling anomaly argument, the power $b$ must be greater than $d+1$.  
Determining the exponent numerically in a reliable manner requires more sophisticated methods for treating the gapless isotropic model, and we leave this as an interesting open question.

\subsection{Effect of anisotropy}
\label{LRO}
Finally, let us numerically study the effect of U(1) symmetry breaking in two and three dimensions by the QMC calculation. To this end, we introduce the anisotropy
\begin{align}
\hat{H}'=-\delta\sum_{\langle\bm{r},\bm{r}'\rangle}(2\hat{s}_{\bm{r}}^x\hat{s}_{\bm{r}'}^x+\hat{s}_{\bm{r}}^y\hat{s}_{\bm{r}'}^y)
\end{align}
and consider the Hamiltonian $\hat{H}_\delta\equiv\hat{H}+\hat{H}'$ just like as we did in the one-dimensional case. 
Even in the presence of $\delta>0$, the Hamiltonian possesses the time-reversal symmetry $\hat{\mathcal{T}}$  and the formation of the N\'eel order requires spontaneous breaking of $\hat{\mathcal{T}}$.  
In principle, we could directly see how the corner magnetization $M_{\text{c}}^z$, defined by Eqs.~\eqref{edgeMhighD} and \eqref{Mczlim1}, gets modified by the anisotropy parameter $\delta$.
However, as we have seen in Sec.~\ref{sec1D} for the 1D case, the deviation of $M_{\text{c}}^z$ from the quantized value is very small and comparable to the statistical error of the QMC results, which makes difficult to estimate it precisely by the direct calculations. 

To overcome this difficulty and estimate $M_{\rm c}^z$ precisely, we consider an imaginary-time correlation of the total magnetization $\hat{S}^{z}_{\rm tot}(\tau)\equiv e^{+\tau\hat{H}_\delta}\hat{S}^{z}_{\rm tot}e^{-\tau\hat{H}_\delta}$:
\begin{align}
C^{zz}_{\rm tot}(\beta)
&\equiv\frac{1}{Z}\text{tr}\left[\hat{S}^{z}_{\rm tot}(\tau=\beta/2)\hat{S}^{z}_{\rm tot}(0)\right]\notag\\
&=\frac{1}{Z}\text{tr}\left[\hat{S}^{z}_{\rm tot}e^{-\frac{1}{2}\beta\hat{H}_\delta}\hat{S}^{z}_{\rm tot}e^{-\frac{1}{2}\beta\hat{H}_\delta}\right],
\label{corrSztot}
\end{align}
where 
$Z=\text{tr}
\, e^{-\beta\hat{H}_\delta}
$ is the partition function. In this approach, we can set $h=0$ from the beginning, just like in the standard treatment of symmetry broken phases by long-range orders.
For a large $\beta$, $C^{zz}_{\text{tot}}(\beta)$ is dominated by 
the doubly-degenerate ground states $|+\rangle$ and $|-\rangle=\hat{\mathcal{T}}|+\rangle$ with well-developed N\'eel order:
\begin{align}
C^{zz}_{\rm tot}(\beta) &\simeq
\frac{1}{2}
\left[ \langle +|\hat{S}^{z}_{\rm tot}|+\rangle^2
+
 \langle -|\hat{S}^{z}_{\rm tot}|-\rangle^2 
 +
2| \langle -|\hat{S}^{z}_{\rm tot}|+\rangle |^2
 \right]\notag
  \\
&=
\frac{1}{2}
\left[ 
\langle +|\hat{S}^{z}_{\rm tot}|+\rangle^2
+
 \langle -|\hat{S}^{z}_{\rm tot}|-\rangle^2 
 \right]\notag\\
 &= \langle +|\hat{S}^{z}_{\rm tot}|+\rangle^2=Q^2.
\end{align}
In going to the second line, we dropped the cross terms, 
which vanishes because $|+\rangle$ and  $|-\rangle$ are linear combinations of the wave function
in different $S^z_{\rm tot}$ sectors, i.e., the sectors of $S^z_{\rm tot} = \pm 1/2 + 2n$ ($n$ is an integer) for $|\pm\rangle$.
Therefore,
$M_{\text{c}}^z$ can be estimated from Eqs.~\eqref{fa2} and \eqref{fa3} as
\begin{align}
M_{\text{c}}^z =\frac{Q}{N}=\frac{1}{2^d}\sqrt{C^{zz}_{\text{tot}}(\beta)}.
\label{mcorner_qmc}
\end{align}
Both $L$ and $\beta$ must be sufficiently large in order to justify all the approximations in the discussions above. As noted in Sec.~\ref{isotropic}, $L$ must be an odd integer to use the filling anomaly formula.
Furthermore, the bulk, the surfaces, and the hinges of the system must be all charge neutral. This is numerically suggested by Figs.~\ref{fig2D_1}(b) and \ref{fig3D_1}(b) but can also be argued in the following way. The neutrality of the bulk follows by the vanishing bulk magnetization, and that of the surfaces can be checked by computing the bulk polarization as we did for the one-dimensional system.  Hinges must also be charge neutral because the three-dimensional system can be understood as stacking of two-dimensional layers with alternating sign of corner magnetizations.

\begin{figure}
\begin{center}
\includegraphics[width=1\columnwidth, trim = 0 400 0 0, clip]{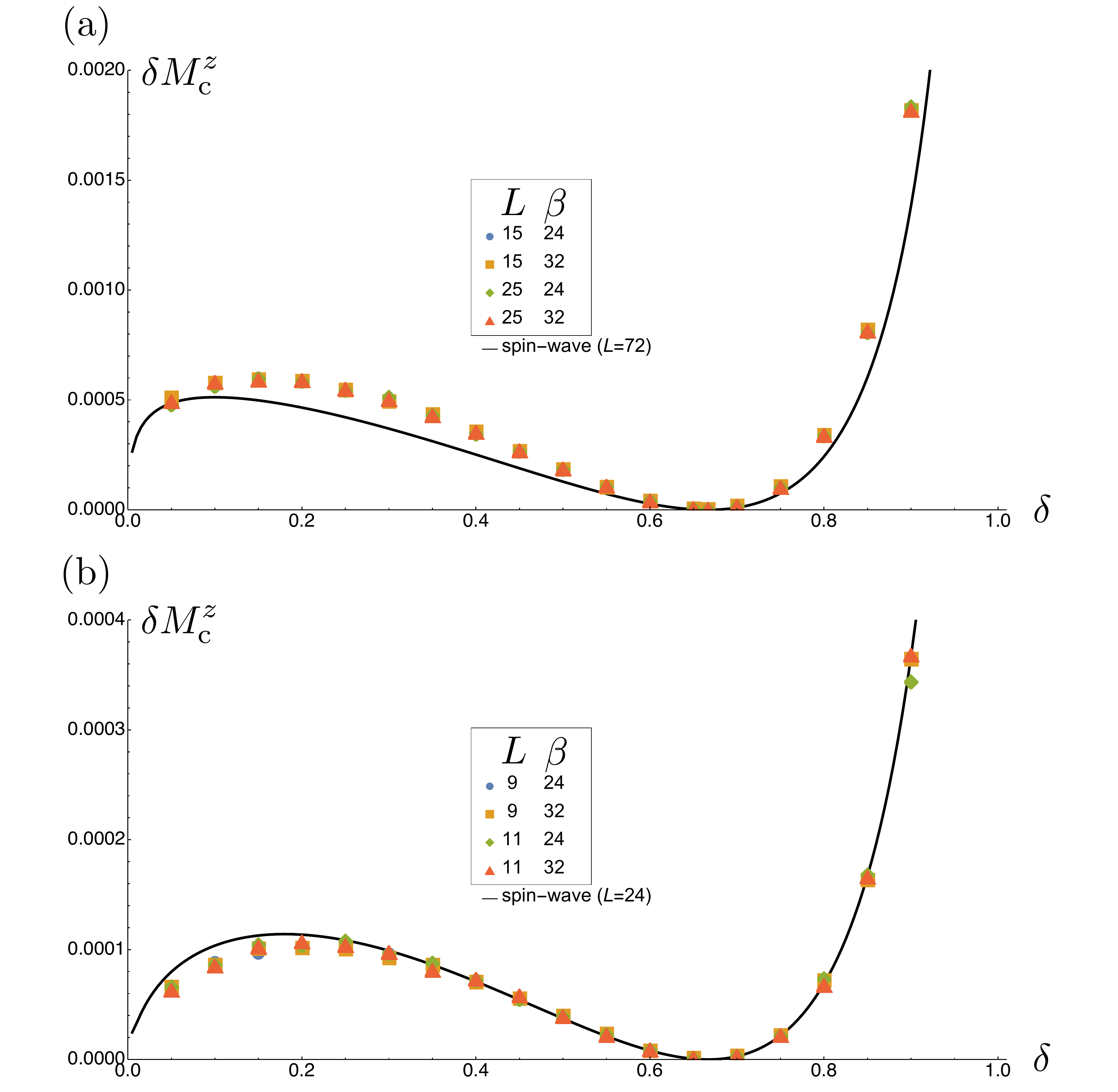}
\caption{
\label{fig2D_2} 
Deviation of the edge magnetization, $\delta M_{\text{c}}^z\equiv S/2^d-M_{\text{c}}^z$, as a function of $\delta$ calculated for the $S=1/2$ Heisenberg model under the OBC defined on (a) the two-dimensional square lattice and (b) the three-dimensional cubic lattice. 
The value of $M_{\text{c}}^z$ is estimated from the correlation function $C_{\text{tot}}^{zz}$ [Eqs.~\eqref{corrSztot} and \eqref{mcorner_qmc}] by the QMC calculations. 
The statistical errors in the QMC data are smaller than the symbol sizes.
The results of the linear spin-wave calculation of  $\delta M_{\text{c}}^z$ ($L=72$ for two dimensions and $L=24$ for three dimensions) are also shown by the black line for comparison.  The staggered field $h$ is set $0$ in these calculations.}
\end{center}
\end{figure}

Our numerical results of $\delta M_{\text{c}}^z \equiv 1/2^{d+1}-M_{\text{c}}^z$ are shown in Fig.~\ref{fig2D_2}. We see that $\delta M_{\text{c}}^z/M_{\text{c}}^z$ is smaller than $10^{-2}$ and the corner magnetization is nearly insensitive to the anisotropy. 
Figures~\ref{fig1Da} and \ref{fig2D_2} clearly suggest the tendency that the QMC calculations and the spin-wave results agree better in higher dimensions: in contrast to the one-dimensional case where the absolute value of $\delta M_{\text{c}}^z $ were underestimated in the spin-wave theory by the factor of $10^2$, the agreement becomes better for higher dimensions and is almost perfect in three dimensions.
These are understood from the fact that quantum fluctuations are less relevant in higher dimensions.

\section{Discussions}
\label{conclusion}
In this work we explored fractional corner magnetizations of collinear antiferromagnets as a direct analog of fractional corner charges in ionic crystals.
We demonstrated that the corner magnetization is quantized to $S/2^d$ in $d$-dimensional cubic system for $d=1$, $2$, and $3$ when the spin rotational symmetry about the $z$ axis is exact. In the presence of an anisotropy which breaks the symmetry, the corner magnetization is no longer quantized but the deviation from the quantized value turns out to be very small (typically about $1$\%), at least for the type of anisotropy considered here. In particular, the $d=2$ and $3$ systems we considered numerically are all gapped because of a finite (though possibly small) staggered field $h$ or anisotropy $\delta$. This allowed us to circumvent the difficulty in studying the magnetically ordered ground state in the finite-temperature simulation by using a sufficiently large inverse temperature $\beta$. Our results suggest that the corner magnetization becomes power-law localized in the isotropic limit due to the emergence of the gapless Goldstone modes, although we leave the detailed analysis of the power-law exponent for future studies.

The fractional corner magnetizations predicted in this work can, in principle, be measured in the actual materials by local probes such as the atomic force microscope. For example, $\mathrm{La}_2\mathrm{Cu}\mathrm{O}_4$, which is a parent material for cuprate high-$T_c$ superconductors, is known to develop a two-dimensional collinear N\'eel order with $S=1/2$ on the Cu square lattice, and hence, would be a good candidate in two dimensions~\cite{PhysRevLett.86.5377,PhysRevLett.105.247001}.
It is interesting to note that recently the compound was successfully fabricated in the form of a single layer~\cite{Dean2012}. 
We note many other candidates for the $S=1/2$ square antiferromagnets, e.g., 
Sr$_2$CuTeO$_6$~\cite{PhysRevLett.117.237203},
MoOPO$_4$~\cite{PhysRevB.96.024445}, and 
Ba$_2$CuTeO$_6$ and Ba$_2$CuWO$_6$~\cite{C8CC09479A}.
For three-dimensional systems, perovskite materials, which develop G-type antiferromagnetic orders, such as $A$TiO$_3$ with $A=$ La, Ce, Pr, Nd, and Sm, would be good candidates~\cite{goodenough1963magnetism,tilley2016perovskites}.
Other materials like TaF$_3$, KFeF$_3$, KCoF$_3$~\cite{goodenough1963magnetism}, and RbMnF$_3$~\cite{PhysRevB.90.054402} would also be worth investigating.

Anisotropy originating from the Dzyaloshinskii--Moriya coupling becomes important near the edges/surfaces of the samples where the inversion symmetry is locally broken. In this work, we could not study such an effect due to the negative sign problem of the QMC calculations. We leave more detailed examination to the future work.

\begin{acknowledgements}
The work of H.W. is supported by JSPS KAKENHI Grant No.~JP20H01825 and by JST PRESTO Grant No.~JPMJPR18LA. 
The work of H.C.P. is supported by a Pappalardo Fellowship at MIT. H.C.P. also thanks the Hong Kong University of Science and Technology for hospitality. 
The work of Y.K. is supported by JSPS Grant Numbers JP18K03447.
The QMC results in the present paper were obtained by the QMC program DSQSS
(https://github.com/issp-center-dev/dsqss).
Numerical calculations were conducted on the supercomputer system in ISSP, The University of Tokyo.
\end{acknowledgements}

\bibliography{ref}

\appendix

\section{Resta's formula}
\label{appResta}
In the main text the polarization $\mathcal{P}$ is computed as the sum of the classical polarization $\mathcal{P}_0$ and the Berry phase correction $\tilde{\mathcal{P}}$.
The polarization can also be computed under the PBC using Resta's formula~\cite{PhysRevLett.80.1800,PhysRevLett.89.077204}. For spin $S$ systems, it reads
\begin{equation}
\mathcal{P}_{\text{R}}
=\begin{cases}
\frac{1}{2\pi}\text{Im}\log\langle e^{i\frac{2\pi}{L}\sum_{r=1}^Lr\hat{s}_r^z}\rangle\text{ mod } 1&(S=0,1,\cdots)\\
\frac{1}{4\pi}\text{Im}\log\langle e^{i\frac{4\pi}{L}\sum_{r=1}^Lr\hat{s}_r^z}\rangle\text{ mod } \frac{1}{2}&(S=\frac{1}{2},\frac{3}{2},\cdots)
\end{cases}
\end{equation}
See Ref.~\cite{PhysRevX.8.021065} for the details of these formulas.

\section{Choice of coarse-graining parameter
\label{app:lambda}}
Here we show that our choice of $\lambda=2$ in this work is practically large enough. 
We first consider an infinite one-dimensional system and assume $\langle \hat{s}_r\rangle=M(-1)^r$ for $r\in\mathbb{N}$. Then the coarse-grained magnetization is given by
\begin{equation}
m^z(r)= M\sum_{r'\in\mathbb{N}}^Lg(r-r')(-1)^{r'}.
\end{equation}
Note that $m^z(r)$ is a periodic function of $r$ satisfying $m^z(r+1)=m^z(r)$. The amplitude of the oscillation of $m^z(r)/M$ is given by
\begin{equation}
A(\lambda)=\frac{m^z(0)}{M}=\sum_{r'\in\mathbb{N}}^Lg(r')(-1)^{r'}=\frac{\vartheta_4(0,e^{-\frac{1}{2\lambda^2}})}{\sqrt{2\pi\lambda^2}},
\end{equation}
where $\vartheta_4(z,q)\equiv\sum_{n\in\mathbb{N}}(-1)^nq^{n^2}e^{2niz}$ is one of Jacobi theta functions. $A(\lambda)$ is monotonically decreasing; for example, 
\begin{align}
A(1)&=1.44\times10^{-2},\\
A(1.5)&=3.01\times10^{-5},\\
A(2)&=5.35\times10^{-9}.
\end{align}
To define the corner magnetization in $d$ dimensional system properly, the coarse-grained magnetization $m^z(\bm{r})$ must be negligibly smaller than $M$ in the bulk, on the surface, and at the hinges. This is guaranteed when $A(\lambda)\ll1$, such as when $\lambda=1.5$ and $\lambda=2$.
\clearpage
\end{document}